\documentclass{article}

\usepackage{amsmath,amssymb,amstext}
\usepackage[dvips]{epsfig}
\usepackage[dvips]{graphicx}
\usepackage{wrapfig}
\usepackage{icrctc07}

\title{Composition-sensitive parameters measured with the surface detector of the Pierre Auger Observatory}
\shorttitle{SD Composition}

\authors{M.D. Healy$^{1}$ for the Pierre Auger Collaboration$^2$}
\shortauthors{Pierre Auger Collaboration}
\afiliations{$^1$University of California, Los Angeles, Los Angeles, CA 90095, USA\\
$^2$Observatorio Pierre Auger, Av. San Mart\'in Norte 304, (5613) Malarg\"ue, Argentina}
\email{healymd@physics.ucla.edu}

\abstract{A key step towards the understanding of the origin of
ultra-high energy cosmic rays is their mass composition.
Primary photons and neutrinos produce markedly different
showers from nuclei, while showers of different nuclear species are
not easy to distinguish. To maximise the discrimination
with the Pierre Auger Observatory ideally all mass-sensitive
observables should be combined, but the 10\% duty
cycle of the fluorescence detector limits the use of direct
measurements of shower maximum at the highest energies.
Therefore, we investigate mass-sensitive observables accessible
with the surface detectors alone.  These are the signal risetime in the
Cherenkov stations, the curvature of the shower front, the
muon-to-electromagnetic ratio, and the azimuthal signal
asymmetry. Risetime and curvature depend mainly on the
depth of the shower development in the atmosphere,
and thus on primary energy and mass. The muon content of
a shower depends on the primary energy and the number of nucleons,
while asymmetry about the shower core is due to geometric effects and
attenuation, which are dependent on the primary mass. The
mass sensitivity of these variables is demonstrated and
their application for composition studies is discussed.}

\begin{document}
\maketitle

\section{Introduction}
\label{sec-intro}

The Pierre Auger Observatory \cite{nimpaper, auger} records air showers
induced by cosmic rays with energies $> 10^{18}$ eV with
unprecedented precision and statistics.
Key to this are the shape of the energy spectrum, the arrival
direction distribution and the composition of the primaries.
Specifically, a transition of galactic to extragalactic cosmic
rays could be indicated by a marked change of composition (see
e.g. \cite{hillas, berezinsky, allard}). While the determination of
the direction and energy of cosmic rays are relatively easy
, the determination of the identity of primary
particles from the details
of the atmospheric extensive
air showers (EAS) is difficult.  Differences between neutrino and
photon induced showers and those initiated by nuclei are relatively
easy to detect \cite{photonlimit1, photonlimit2, neutrinolimit}.
%
%
The differences between nuclear species (e.g. proton and iron) are more
difficult to measure: on average heavier nuclei (of the same total
energy) develop higher in the atmosphere (lower $X_{\rm max}$), due to
their larger cross-sections and the lower energy per nucleon, 
produce more secondary particles, and have a higher 
muon-to-electron ratio at
observation level.  These differences are small and easily concealed
by the variations due to the primary energy or by uncertainties in
hadronic interaction models, and are covered by shower fluctuations
and the limited precision of the measurement with a realistic
detector.  The Pierre Auger Observatory is a hybrid detector, combining a large array of
water-Cherenkov detectors (surface detector, SD, with 100\% duty
cycle) and 24 fluorescence telescopes (FD, with 10\% duty
cycle). Hybrid events have a good energy estimate and provide a direct
measurement of $X_{\rm max}$. But 90\% of the events have information
only from the surface array and, therefore, observables from the array are
needed for composition analysis at the highest energies.
In the following we discuss four observables which all rely on the time
structure of the shower front as measured with a 25 ns flash ADC system
(FADC) of the water-Cherenkov detectors.  They are all sensitive to
the primary mass, via their dependence on the height of shower
development and the muon content in the shower: ({\bf i}) The signal 
risetime in water tanks at some distance from the shower core depends
mainly on the height of the shower development: the higher the shower
development in the atmosphere, the smaller the time spread between the
particles and the smaller the risetime.  ({\bf ii}) The curvature of the
shower front reflects the height of the shower development: the
further the main shower development from the detector, the flatter
the shower front.  ({\bf iii}) The muon content of a shower can be
assessed by looking for tell tale signs of sudden, large energy
deposits in the time traces, as they are expected from muons:  the
frequency of large jumps in the signal from one FADC bin to the next
relates to the abundance of muons.  ({\bf iv}) The azimuthal asymmetry in
the signal risetimes within the shower plane is an effect of a
varying muon-to-electromagnetic component, and as such it depends on
the shower development and the muon content.

While all of these variables could in principle be used for composition
studies with a perfect detector, there are limitations due to the
finite accuracy with which they can be measured.  Therefore,
it needs to be shown that measurement uncertainties are small
enough that the mass differences can be detected reliably.
Once the mass sensitivity of the individual variables is established,
a multivariate analysis method will be needed to combine all
mass-sensitive observables to maximise the mass separation power. A
combination of risetime and curvature has already been used
successfully to pose an upper limit on primary photons
\cite{photonlimit2}.
\section{The risetime of the signal, t$_{1/2}$}
\label{sec-risetime}
The signals in the 10 m$^2$ water-Cherenkov detectors are
characterised by $t_{1/2}$, the time to rise from 10 to 50\% of the
integrated signal. The signals, up to 3 $\mu$s long, are recorded
with a 25 ns FADC system. The risetime in an individual station depends
on core distance, zenith angle and energy and has been parameterised from
experimental data to correct events from different zenith angles.
\begin{figure}[b]
\begin{center}
\epsfig{file=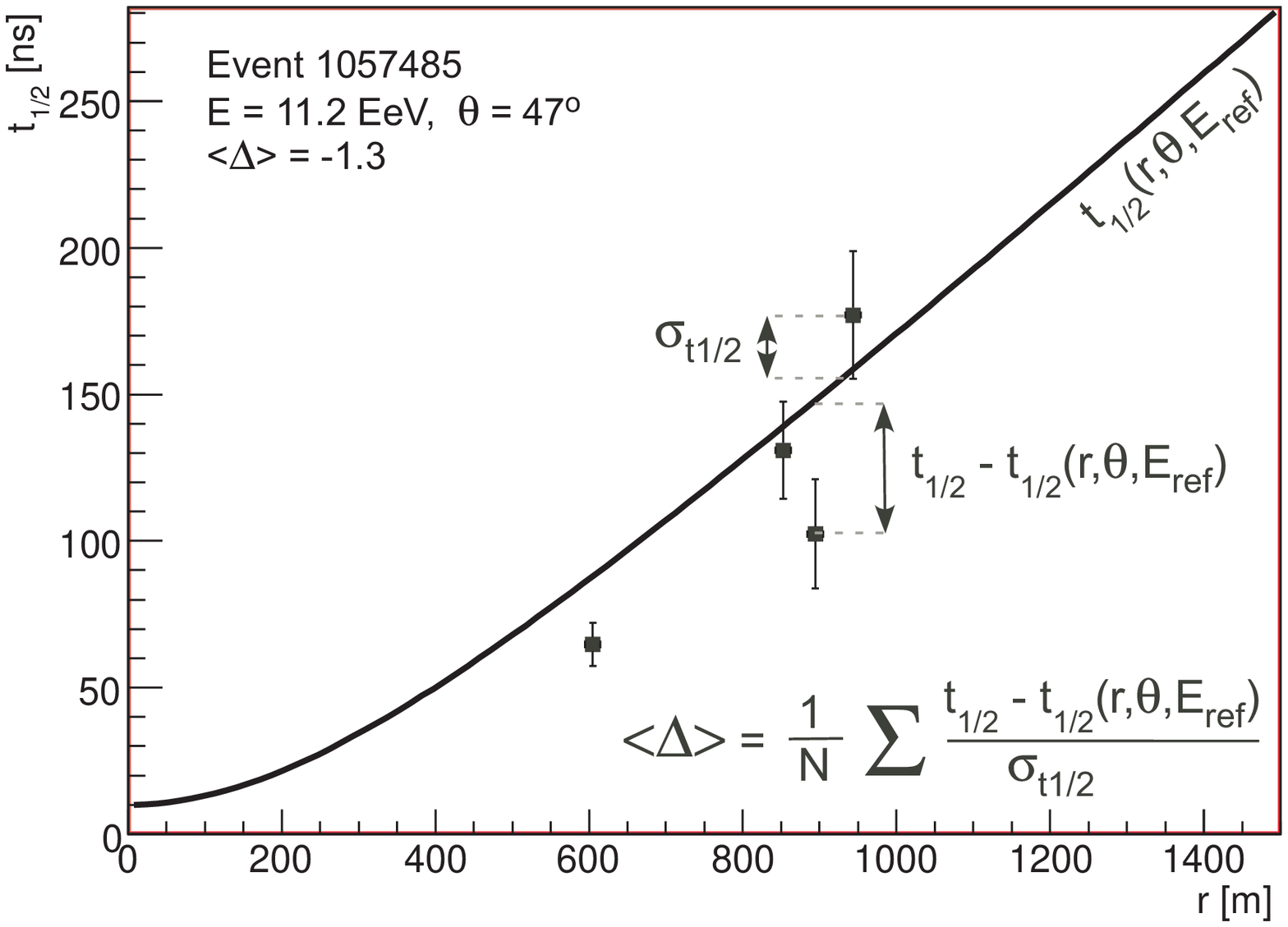,width=72.0mm}\\
\epsfig{file=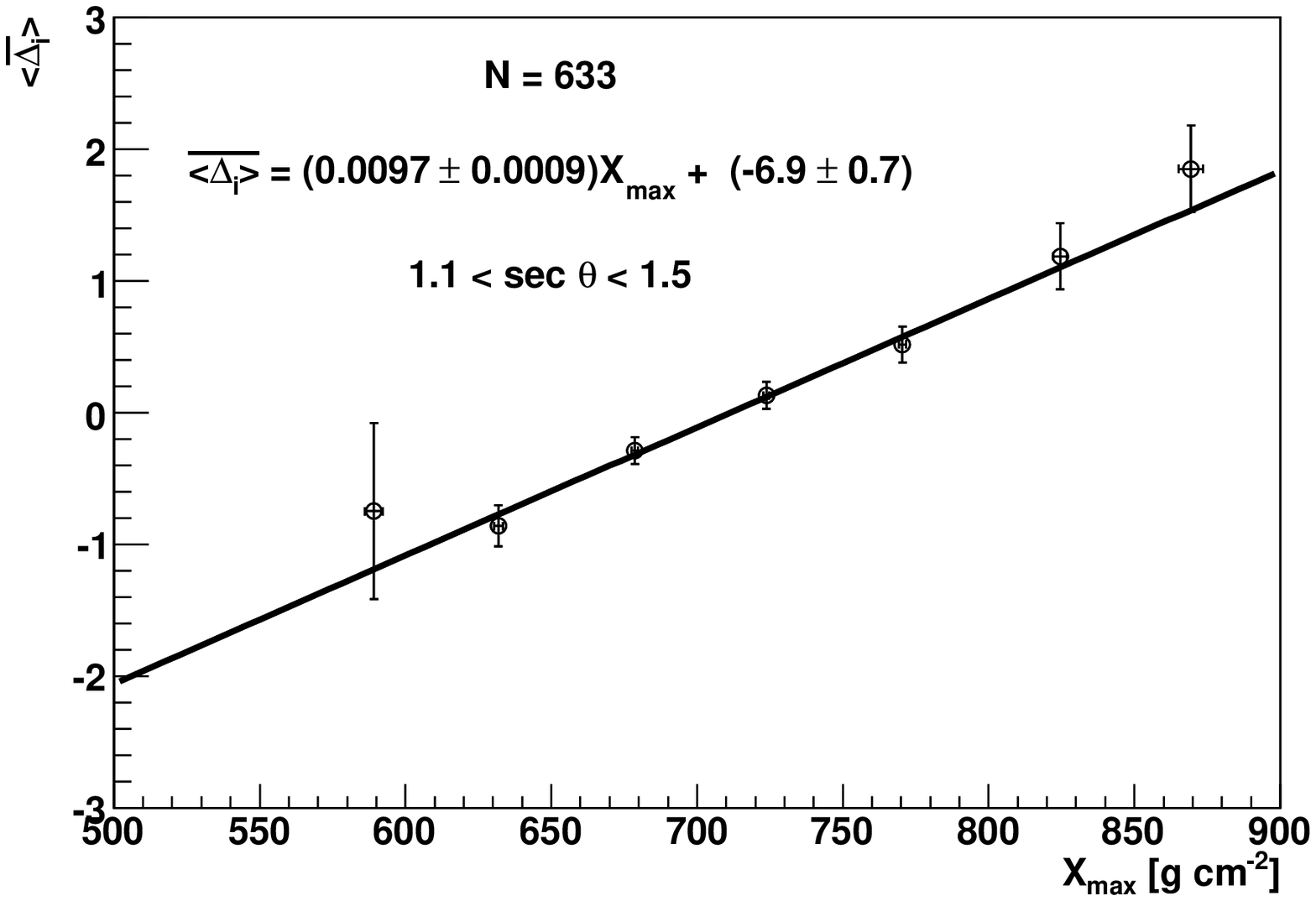,width=72.0mm}
\end{center}
\caption{{\bf Top:} $\langle\Delta\rangle$ derived from a
  single event.  The black line is the predicted risetime while the data
  points represent measurements with the error bars from
  twin station studies.  Combining each stations' $\Delta$ from the
  expectation an average for the
  event can be defined.  {\bf Bottom:} $\overline{\langle\Delta\rangle}$
  as a function of $X_{\rm max}$, for nearly-vertical hybrid events.}
\label{fig-rise}
\end{figure}
The measurement uncertainty $\sigma_{t_{1/2}}$ as function of signal
$S$ in the tank, the core distance $r$ and zenith angle $\theta$ has
been determined from a small number of twin tanks and from pairs of
detectors at similar core distances (within 100 m). A
parameterisation of $\sigma_{t_{1/2}}$ is used to define the error bars
shown in fig. \ref{fig-rise} (top). Measurements of $t_{1/2}$ for
signals $< 20$ VEM are severely influenced by Poisson fluctuations due
to low numbers of particles and are not used. A correction of the
risetime for the azimuthal asymmetry in the shower, 
particularly important between 35 and 50$^\circ$, is applied.
For each station the deviation $\Delta$ of the measured risetime, in units
of the accuracy of measurement, from the average values found for $r$
and $\theta$ at a fixed reference energy $E \approx 10^{19}$ eV is
formed, thus: $ \Delta = \frac{t_{1/2}(r, \theta) - \langle t_{1/2}(r,
\theta, E_{\rm ref})\rangle}{\sigma_{t_{1/2}}(S,r,\theta)} $.  All stations
in one event are then combined in an average event deviation
$\langle\Delta\rangle$ as seen in fig. \ref{fig-rise} (top), which should be
larger for showers developing deeper than the reference and smaller for
those developing higher.
Using hybrid events \cite{hybrid} it can be tested whether the
so-constructed variable $\langle\Delta\rangle$ indeed correlates with
$X_{\rm max}$.  The data have been binned in $X_{\rm max}$ 
, and for all events in one $X_{\rm max}$ bin
the average of $\langle\Delta\rangle$ (i.e.
$\overline{\langle\Delta\rangle}$) has been formed.  This is shown in
fig. \ref{fig-rise} (bottom).  A linear dependence is observed which
allows estimation of $X_{\rm max}$ from events observed with the SD alone.
%
%
\section{Shower Front Curvature}
\label{sec-curvature}
\begin{figure}[ht]
\begin{center}
\epsfig{file=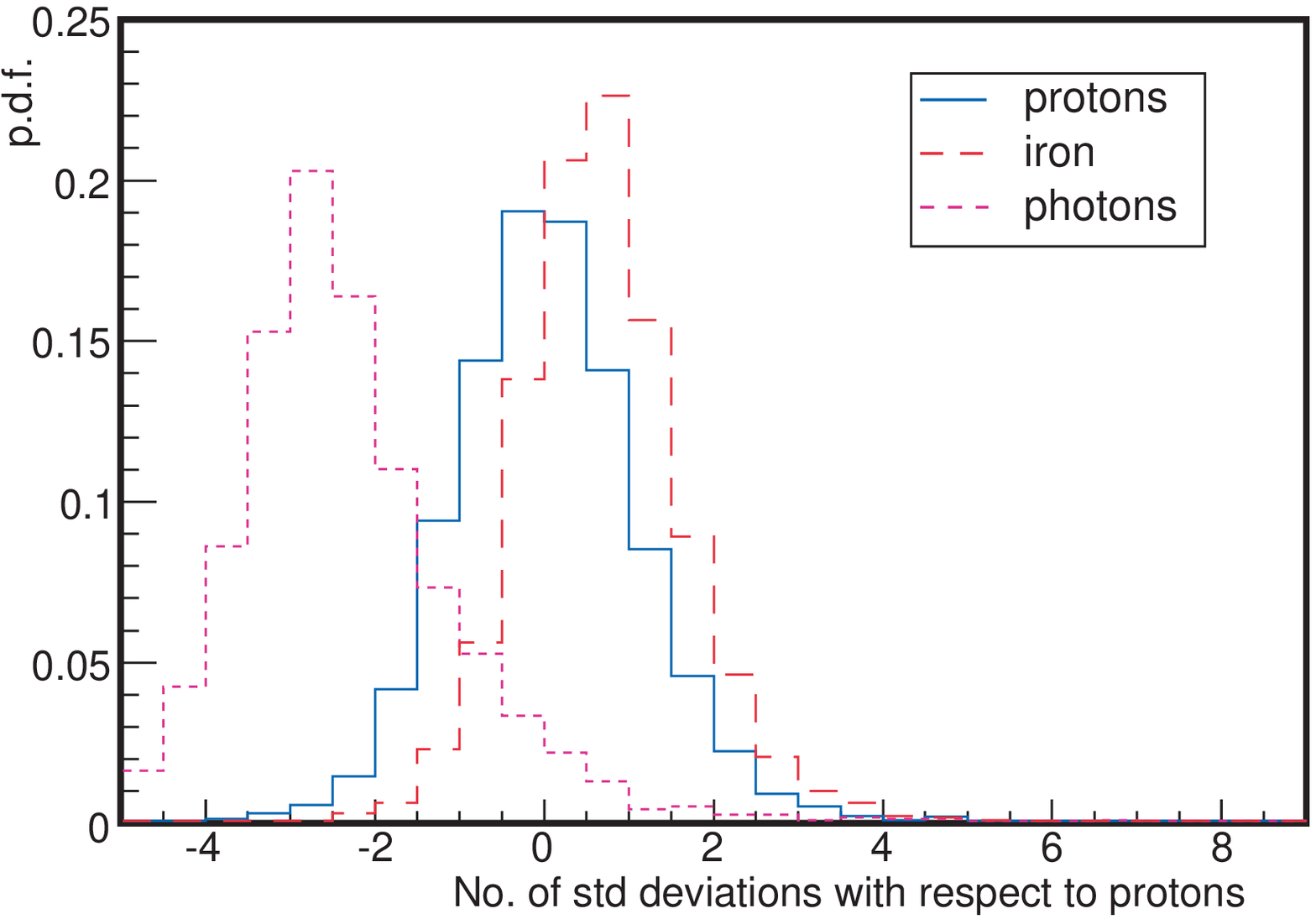,width=66.0mm}\\
\epsfig{file=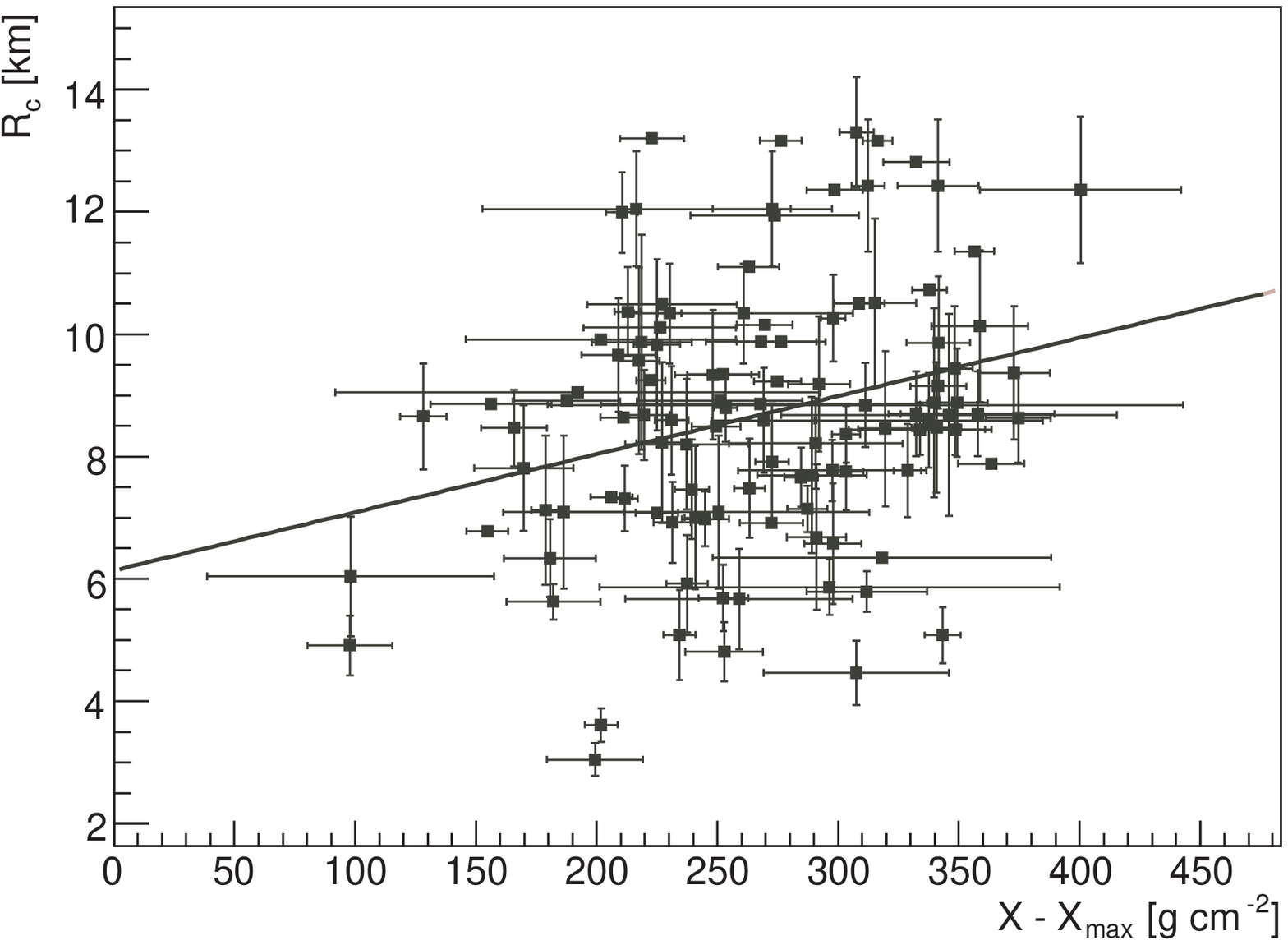,width=66.0mm}
\end{center}
\caption{{\bf Top:} Comparative radii of curvature from simulations
  of three primaries (solid-blue: proton, solid-red: iron, dashed-magenta: 
  photon).  {\bf Bottom:} 
  ${\rm R_c}$ is shown as a function of $X-X_{\rm max}$ for
  nearly vertical hybrid showers ($1.1<\sec(\theta)<1.2$).  The line
  is a linear fit to the data points correlating ${\rm R_c}$ with the distance
  from the detector to the shower maximum.}
\label{Hybrid_Xmax_Curve}
\end{figure}

The trigger times of the array detectors away from the core lag behind
what one would expect from a plane of particles traveling at the speed
of light, indicating the shower front is curved.  With the assumption
of a hemispherical shower front
a {\em radius of curvature}, R$_c$, can be derived.  The radius of curvature
is an SD observable that Monte Carlo simulations show is mass sensitive (see 
Fig. \ref{Hybrid_Xmax_Curve}).
Larger
distances of $X_{\rm max}$ from the array yield larger radii of curvature,
while deeply penetrating showers produce shower fronts with smaller
radii of curvature.
Also muons tend to travel in straight lines and electrons are more
scattered, so muons arrive at the detectors first, and with a narrower
time spread, as compared to the electro-magnetic component.
Therefore, a shower with a larger number of muons will produce
station triggers that are compact in time, resulting in a flatter 
shower front than for a similar shower that is muon-poor.
The radius of curvature is strongly dependent on zenith angle because
the distance from the array to shower maximum increases with zenith angle ($\propto (X_0\sec\theta -X_{\rm max})$).  Fig.~\ref{Hybrid_Xmax_Curve}
demonstrates the typical resolution between the radius of curvature
and changes in $X-X_{\rm max}$.
%
\section{Muon Content}
\label{sec-muons}
\begin{figure}[t]
\begin{center}
  \epsfig{file=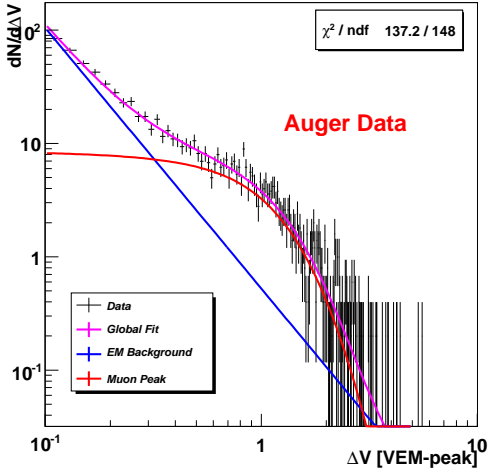,width=69.0mm}\hfill
\end{center}
\caption{Distribution of jumps for Auger events 
  with energy between 8-12 EeV and 40-50$^\circ$, averaged over 251
  stations - 100 showers - within the same core distance
  interval.
}
\label{fig-jump-dist}
\end{figure}

As a typical muon deposits much more energy ($\approx$ 240 MeV) in a
water tank than an electron or photon ($\approx$ 10 MeV), 
spikes are produced over the smoother electromagnetic background in the
FADC time traces. Thus, muons manifest themselves as sudden variations
in the signal $\Delta V$ from one FADC bin to the next.  The expected
distributions of $\Delta V$ for purely electromagnetic and muonic
traces look very different. A mix of electromagnetic and
muonic signals is needed to fit the measured distribution (see
Fig.~\ref{fig-jump-dist}).
The number of muons in the shower is then estimated via the number of $\Delta V$s
above 0.5 VEM with $N_\mu = \alpha N(\Delta V > 0.5 {\rm VEM})$; with $\alpha
\approx 1.4$ being only weakly dependent on core distance and zenith
angle. The $N_\mu$ resolution is about 25\%, systematic errors are 
below 10\%.
\section{Azimuthal Asymmetry}
\label{sec-asymm}
The observed azimuthal asymmetry in the risetimes offers another 
handle on primary composition determination.
Its magnitude is a measure of the degree of shower development 
and hence it is sensitive to primary composition \cite{tere-asym}.
The dependence of the risetimes with azimuthal angle $\zeta$
is fitted using $t_{1/2}(r,\zeta) = a + b\,\cos\zeta$, for all stations 
with 500 m $ < r < $ 2000 m and with signals greater than 10 VEM.
\begin{figure}[t]
\begin{center}
  \epsfig{file=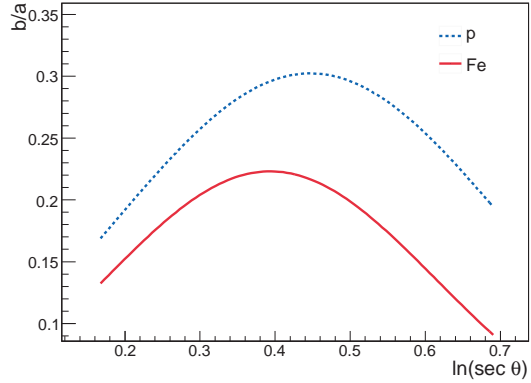,width=69.0mm}
\end{center}
\caption{Development of the asymmetry factor in the risetime with 
  zenith angle. The primary energy is $\sim 10^{18.5}eV$.}
\label{riseasym}
\end{figure}
The asymmetry factor $\frac{b}{a}$ is sensitive to the distance between
the detector and the shower maximum (shown in fig. \ref{riseasym} as 
$\sec\theta \propto X-X_{\rm max}$).  Maximal asymmetry occurs at a
unique value of $X-X_{\rm max}$ found most frequently at a corresponding
zenith angle. Monte Carlo simulations show that this zenith angle ($\theta_{asymax}$)
is different for proton and iron showers (Fig. \ref{riseasym}) reflecting the 
systematic difference in the average $X_{\rm max}$.
%
\bibliography{icrc0596}

\begin{thebibliography}{10}

\bibitem{photonlimit2}
J.~{Abraham et al. [Pierre Auger Collaboration]}.
\newblock {\em Submitted to Astrop. Phys.}

\bibitem{nimpaper}
J.~{Abraham et al. [Pierre Auger Collaboration]}.
\newblock {\em Nucl. Instr. Meth. A}, 523:50--95, 2004.

\bibitem{photonlimit1}
J.~{Abraham et al. [Pierre Auger Collaboration]}.
\newblock {\em Astrop. Phys.}, 27:155, 2007.

\bibitem{allard}
D.~{Allard et al.}
\newblock {\em Astrop. Phys.}, 27:61, 2007.

\bibitem{berezinsky}
V.~{Berezinsky et al.}
\newblock {\em Phys. Rev. D.}, 74:61, 2006.

\bibitem{neutrinolimit}
O.~{Blanch [Pierre Auger Collaboration]}.
\newblock In {\em {\rm these proceedings (\#603)}}.

\bibitem{tere-asym}
M.~T. {Dova [Pierre Auger Collaboration]}.
\newblock In {\em {\rm Proc. 29\textsuperscript{th} ICRC (Pune)}}, pages
  101--104, 2005.

\bibitem{hillas}
A.~M. {Hillas}.
\newblock {\em J. Phys. G}, 31:95, 2005.

\bibitem{auger}
T.~{Suomijarvi [Pierre Auger Collaboration]}.
\newblock In {\em {\rm these proceedings (\#299)}}.

\bibitem{hybrid}
M.~{Unger [Pierre Auger Collaboration]}.
\newblock In {\em {\rm these proceedings (\#594)}}.

\end{thebibliography}
\bibliographystyle{plain}
\clearpage
\end{document}